\documentclass[12pt]{article}
\usepackage{amsmath}
\usepackage{graphicx}
\begin{document}
\baselineskip=18 pt
\begin{center}
{\large{\bf  Cylindrically symmetric, Asymptotically flat, Petrov Type D spacetime with a Naked Curvature Singularity and Matter Collapse}}
\end{center}

\vspace{.5cm}

\begin{center}
{\bf Faizuddin Ahmed}\footnote{faizuddinahmed15@gmail.com,{\\{Permanent address :
{\it Hindustani Kendriya Vidyalaya, Dinesh Ojha Road,}\\
{\it Guwahati-781005, India}}}}\\
{\it Department of Physics, Cotton College,}\\
{\it Guwahati-781001, India}
\end{center}

\vspace{.5cm}

\begin{abstract}

We present a cylindrically symmetric, Petrov type D, nonexpanding, shear-free and vorticity-free solution of Einstein's field equations. The spacetime is asymptotically flat radially and regular everywhere except on the symmetry axis where it possesses a naked curvature singularity. The energy-momentum tensor of the spacetime is that for an anisotropic fluid which satisfies the different energy conditions. This spacetime is used to generate a rotating spacetime which admits closed timelike curves and may represent a Cosmic Time Machine.

\end{abstract}

\vspace{0.1cm}

\vspace{.5cm}

\section{Introduction}

A spacetime is said to have singularities if it possesses inextendible curves with finite affine length. The singularities of spherically symmetric matter filled spacetimes can be recognized from the divergence of the energy density and curvature scalars, for example, the Kretschmann scalar $R_{\mu\nu\rho\sigma}\,R^{\mu\nu\rho\sigma}$ and $R_{\mu\nu\rho\sigma}\,R^{\rho\sigma\lambda\tau}\,R^{\mu\nu}_{\lambda\tau}$. These singularities are of two kinds, spacelike singularity (e.g., Schwarzschild singularity) and timelike singularity. In timelike singularities, two possibilities arise (i) There are an even number of horizons around a timelike singularity (e.g., RN black hole). In that case, an observer cannot see this singularity from outside (the singularity is covered by an event horizon). (ii) There is no horizon around a timelike singularity and the later one may be observable from the rest of the spacetime.

To counter the occurence of timelike singularities which are not covered by an event horizon, R. Penrose \cite{Pen1,Pen2,Pen3} proposed a Conjecture known as Cosmic Censorship Conjecture. The Cosmic Censorship Conjecture (CCC) essentially states that no physically realistic collapse, evolving from a well-posed initial data set and satisfying the dominant energy condition, results in a singularity in the causal past of null infinity. According to Weak Cosmic Censorship, singularities have no effect on distant observers, that is, they cannot communicate with far-away observers. Till date, there are no mathematical theorems or proofs that support (or counter) this Conjecture. On the contrary, there is no mathematical reason that a naked singularity cannot exist in a solution of the field equations in GR. In literature, there are a number of examples of gravitational collapse that form a naked curvature singularity. The earliest model is the Lemaitre-Tolman-Bondi (LTB) \cite{Lei,Tol,Bondi} solution, a spherically symmetric inhomogeneous collapse of dust fluid that admits both naked and covered singularities. A small sample of gravitational collapse with the formation of a naked singularity is discussed in \cite{pap,vai,Christ,Jos1,Jos2,Jos3,Gov,Brave,sing1,sing4}. In this paper, we construct a nonvacuum metric that forms a naked curvature singularity. The presence of a naked singularity in a solution of Einstein's field equations may breakdown the causality condition and may form closed timelike curves (CTCs). The possibility that a generic naked curvature singularity gives rise to a Cosmic Time Machine has been discussed by Clarke and de Felice \cite{Clark} (see also \cite{Felice,Felice2,Felice3}). A Cosmic Time Machine is a spacetime which is asymptotically flat and admits closed non-spacelike curves which extend to future infinity. In literature, there are a number of vacuum as well as non-vacuum spacetimes admitting CTCs have been constructed without naked singularity. Small samples of these are in \cite{Go,Sto,Tip,Gott,Ori1,Ori2,Bon,Bon1,Bon2,Taub,Kerr,Sar,Faiz}. Recently, a vacuum spacetime with a naked curvature singularity admitting CTCs appeared in \cite{Sar2} which may represent a Cosmic Time Machine. One way of classifying causality violating spacetimes would be to categorize the metrics as either eternal time machines in which CTCs always exist (in this class would be \cite{Go,Sto}) or as time machine spacetimes in which CTCs appear after a certain instant of time (in this class would be \cite{Ori1,Ori2}). However, many of the spacetimes admitting CTC suffer from one or more severe physical problems. The time machine spacetime discussed in \cite{MTY,Morri,Alcu,Everett,Eve} violated the weak energy condition (WEC) and the strong energy condition (SEC) is violated in \cite{Ori3,KD,Ori4,Ori5}. The WEC states that for any physical (timelike) observer the energy density is nonnegative, which is the case for all known types of (classical) matter fields. Hawking proposed a Chronology Protection Conjecture \cite{Haw} which states that the laws of physics will always prevent a spacetime from forming closed timelike curves (CTCs). But the general proof of the Chronology Protection Conjecture has not yet existed.

\section{Analysis of the Spacetime} 
  
Consider the following cylindrically symmetric metric in $(t,r,\phi,z)$ coordinates:
\begin{equation}
ds^2=\sinh^{2} 2\,r\,dr^2+dz^2+\sinh^{2} r\,\left(d\phi^2-dt^2\right),
\label{metric}
\end{equation}
where the units are chosen such that $c=1$ and $8\,\pi\,G=1$. The coordinates are labelled as $x^0=t$, $x^1=r$, $x^2=\phi$ and $x^3=z$. The ranges of the coordinates are $0 \leq r < \infty$, $-\infty < z < \infty$, $-\infty < t < \infty$ and $\phi$ is periodic coordinate, $\phi\in[0, 2\,\pi)$. The metric is Lorentzian with signature $(-,+,+,+)$ and the determinant of the corresponding metric tensor $g_{\mu\nu}$ is
\begin{equation}
det\;g=-\sinh^{4} r\,\sinh^{4} 2\,r.
\label{det}
\end{equation}
The nonzero components of the Einstein tensor are
\begin{eqnarray}
G^{t}_{t}&=&G^{\phi}_{\phi}=G^{z}_{z}=-\frac{1}{4\,\sinh^{4} r}\\
\label{Eins}
G^{r}_{r}&=&\frac{1}{4\,\sinh^{4} r}\nonumber.
\end{eqnarray}
The scalar curvature invariants are given by
\begin{eqnarray}
R^{\mu}_{\,\,\mu}&=&R=\frac{1}{2\,\sinh^{4} r},\nonumber\\
R_{\mu\nu}\,R^{\mu\nu}&=&\frac{1}{4\,\sinh^{8} r},\\
R_{\mu\nu\rho\sigma}\,R^{\mu\nu\rho\sigma}&=&\frac{3}{4\,\sinh^{8} r},\nonumber\\
R_{\mu\nu\rho\sigma;\tau}\,R^{\mu\nu\rho\sigma;\tau}&=&\frac{4}{\sinh^{12} r}\nonumber.
\label{scalar}
\end{eqnarray}
These curvature scalars and first order differential invariants blow up (or diverge) at $r=0$, indicating the existence of a naked curvature singularity. Since the singularity occurs without an event horizon, the Cosmic Censorship Conjecture has no physical interest. These curvature scalars vanish rapidly at spatial infinity, that is, $r\rightarrow \infty$, indicating that the spacetime is asymptotically flat radially.  

\section{Cylindrical Symmetry and Petrov Type of the Spacetime}

The spacetime represented by (\ref{metric}) is cylindrically symmetric as it is clear from the following. Consider the Killing vector ${\bf \eta}=\partial_{\phi}$ which has the normal form
\begin{equation}
\eta^{\mu}=\left(0,0,1,0\right ).
\label{kill1}
\end{equation}
Its covector form is
\begin{equation}
\eta_{\mu}=\sinh^{2} r\,\left(0,0,1,0\right ).
\label{kill2}
\end{equation}
The vector (\ref{kill1}) satifies the Killing equation $\eta_{\mu\,;\,\nu}+\eta_{\nu\,;\,\mu}=0$. The spacetime (\ref{metric}) that has an axially symmetric axis is assured by the condition \cite{Carter,Marc,Carot,Wang,Steph}
\begin{equation}
\boldsymbol{X}\equiv||\partial_{\phi}||^{2}=|g_{\mu\nu}\,\eta^{\mu}\,\eta^{\nu}|=|g_{\phi\phi}|=\sinh^{2} r\rightarrow 0,
\label{kill3}
\end{equation}
as $r\rightarrow 0^{+}$, where we have chosen the radial coordinate $r$ such that the symmetry axis is located at $r=0$. In addition to this, the spacetime admit another spacelike Killing vector ${\bf \xi}=\partial_{z}$ which, with the axial Killing vector ${\bf \eta}=\partial_{\phi}$ generates a two-dimensional isometry group and they commute. Thus the isometry group is Abelian and the presented spacetime is cylindrically symmetric \cite{Marc,Carot}. However, the spacetime is not regular near to the axis as it fails to satisfy the regularity condition (or elementary flatness condition) \cite{Marc,Carot,Wang,Steph}, namely,
\begin{equation}
\frac{(\nabla_{\mu}\boldsymbol{X})\,(\nabla^{\mu}\boldsymbol{X})}{4\,\boldsymbol{X}}\rightarrow 1,
\label{kill4}
\end{equation}
in the limit at the rotation axis, that is, $r \rightarrow 0$. The appearance of spacetime singularities on the axis can be considered as representing the existence of some kind of sources \cite{Steph,BB,BB1,Sil,Sil1}.

For the classification of the metric (\ref{metric}), we construct the following set of a null tetrad $(k,l,m,\bar{m})$ and they are
\begin{eqnarray}
\label{n1}
k_{\mu}&=&\frac{\sinh r}{\sqrt{2}}\,\left (1,0,1,0\right ),\nonumber\\
\label{n2}
l_{\mu}&=&\frac{\sinh r}{\sqrt{2}}\,\left(1,0,-1,0\right ),\\  
\label{n3}
m_{\mu}&=&\frac{1}{\sqrt{2}}\,\left(0,\sinh 2\,r,0,i\right ),\nonumber\\ 
\label{n4}
\bar{m}_{\mu}&=&\frac{1}{\sqrt{2}}\,\left(0,\sinh 2\,r,0,-i\right ),\nonumber
\end{eqnarray}
where $i=\sqrt{-1}$. The set of null tetrads above are such that the metric tensor for the line element (\ref{metric}) can be expressed as
\begin{equation}
g_{\mu \nu}=-k_{\mu}\,l_{\nu}-l_{\mu}\,k_{\nu}+m_{\mu}\,\bar{m}_{\nu}+\bar{m}_{\mu}\,m_{\nu}.
\label{n5}
\end{equation}
The vectors (\ref{n1})--(\ref{n4}) are null vectors and are orthogonal except for $k^{\mu}\,l_{\mu}=-1$ and $m_{\mu}\,{\bar m}^{\mu}=1$. Using this null tetrad above, we have calculated the five Weyl scalars, of which only
\begin{equation}
\Psi_2=\frac{1}{12\,\sinh^{4} r}
\label{n6}
\end{equation}
is nonvanishing, while $\Psi_0=\Psi_1=0=\Psi_3=\Psi_4$. The metric is clearly of type D in the Petrov classification scheme. The nonvanishing Weyl scalar $\Psi_2$ vanishes rapidly at spatial infinity along the radial direction, that is, $r\rightarrow \infty$, again indicating the asymptotic flatness. 

\section{Matter Field Distribution of the Spacetime and Kinematic Parameters}

For spacetime (\ref{metric}), we consider the energy-momentum tensor for an anisotropic fluid given by
\begin{equation}
T_{\mu\nu}=(\rho+p_{t})\,U_{\mu}\,U_{\nu}+p_{t}\,g_{\mu\nu}+(p_{r}-p_{t})\,\zeta_{\mu}\,\zeta_{\nu},
\label{EMT}
\end{equation}
where $U^{\mu}$ is the unit timelike four-velocity vector and $\zeta_{\mu}$ is the unit spacelike vector along the radial direction $r$. Here the physical parameters $\rho$, $p_r$ and $p_t$ are the energy density, the radial pressure and the tangential pressure of the ansiotropic fluid, respectively.

For metric (\ref{metric}), the timelike unit four-velocity vector $U^{\mu}$ is defined by 
\begin{eqnarray}
U^{\mu}&=&(\frac{1}{\sinh r},0,0,0),\nonumber\\
U^{\mu}\,U_{\mu}&=&-1.
\label{time-like}
\end{eqnarray}
And the spacelike unit vector $\zeta_{\mu}$ along the radial direction is
\begin{eqnarray}
\zeta_{\mu}&=&\left(0,\sinh 2\,r,0,0\right),\nonumber\\
\zeta_{\mu}\,\zeta^{\mu}&=&1.
\label{radial-vector}
\end{eqnarray}
The nonzero components of the energy-momentum tensor (\ref{EMT}) using (\ref{time-like}) and (\ref{radial-vector}) are
\begin{eqnarray}
T^{t}_{t}&=&-\rho,\nonumber\\
T^{r}_{r}&=&p_{r},\nonumber\\
T^{\phi}_{\phi}&=&T^{z}_{z}=p_{t},
\label{energy-com}
\end{eqnarray}
and the trace of the energy-momentum tensor (\ref{EMT}) is
\begin{equation}
T^{\mu}_{\,\mu}=T=p_{r}+2\,p_{t}-\rho.
\label{trace}
\end{equation}
Einstein's field equations (taking cosmological constant $\Lambda=0$) are given by
\begin{equation}
G^{\mu\nu}=T^{\mu\nu},\quad \mu,\nu=0,1,2,3,
\label{Einstein}
\end{equation}
where $G^{\mu\nu}$ is the Einstein tensor and $T^{\mu\nu}$ is the energy-momentum tensor. Thus we have from the field equations (\ref{Einstein}) using (\ref{Eins}), (\ref{energy-com}), and (\ref{trace})
\begin{eqnarray}
\rho&=&\frac{1}{4\,\sinh^{4} r},\nonumber\\
p_{r}&=&\frac{1}{4\,\sinh^{4} r},\\
\label{comp-ener-tensor}
p_{t}&=&-\frac{1}{4\,\sinh^{4} r},\nonumber\\
T&=&-R,\nonumber
\end{eqnarray}
which are diverging at $r=0$, a fact that indicates the collapse of anisotropic fluid on the symmetry axis. The matter distribution satisfies the different energy conditions \cite{Hak}:
\begin{eqnarray}
WEC\,&:& \quad \rho>0,\nonumber \\
WEC_{t}\,&:&\quad  \rho+p_{t}=0,\nonumber \\
WEC_{r}\,&:&\quad  \rho+p_{r}>0,\\
\label{con}
SEC\,&:&\quad \rho+p_{r}+2\,p_{t}=0,\nonumber \\
DEC\,&:&\quad \rho=|p_{j}|\nonumber,\quad j=t,r.
\end{eqnarray}
The kinematic parameters, the {\it expansion} $\Theta$, the {\it acceleration} vector $\dot{U}^{\mu}$, the {\it shear tensor} $\sigma_{\mu\nu}$ and the {\it vorticity tensor} $\omega_{\mu\nu}$ associated with the fluid four-velocity vector are defined by
\begin{eqnarray}
\Theta&=&U^{\mu}_{\,\,;\,\mu},\nonumber\\
a^{\mu}&=&\dot{U}^{\mu}=U^{\mu\,;\,\nu}\,U_{\nu},\nonumber\\
\sigma_{\mu\nu}&=&U_{(\mu\,;\,\nu)}+\dot{U}_{(\mu}\,U_{\nu)}-\frac{1}{3}\,\Theta\,h_{\mu\nu},\\
\label{parameters}
\omega_{\mu\nu}&=&U_{[\mu\,;\,\nu]}+\dot{U}_{[\mu}\,U_{\nu]},\nonumber
\end{eqnarray}
where $h_{\mu\nu}=g_{\mu\nu}+U_{\mu}\,U_{\nu}$ is the projection tensor. For spacetime (\ref{metric}), these parameters have the following expressions:
\begin{eqnarray}
\Theta&=&0=\sigma_{\mu\nu}=\omega_{\mu\nu},\\
\label{para}
a_{\mu}&=&\mbox{coth} r\,\delta^{r}_{\mu}\nonumber.
\end{eqnarray}
The magnitude of the acceleration vector $a^2=a^{\mu}\,a_{\mu}$ is $a=\frac{1}{2}\,\mbox{csch}^2 r$.

\section{Generating a Rotating Spacetime Admitting CTCs : A Cosmic Time Machine}

One can easily generate a rotating spacetime mixing of the $dT\,d\psi$ term from metric (\ref{metric}) by the following transformations, keeping $r$ and $z$ the same:
\begin{eqnarray}
t&\rightarrow& -e^{-\frac{\psi}{2}}+T\,e^{\frac{\psi}{2}},\nonumber\\
\phi&\rightarrow& e^{-\frac{\psi}{2}}+T\,e^{\frac{\psi}{2}}.
\label{trans}
\end{eqnarray}
From metric (\ref{metric}), we get 
\begin{equation}
ds^2=\sinh^{2} 2\,r\,dr^2+dz^2-\sinh^{2} r\,(T\,d\psi^2+2\,dT\,d\psi ),
\label{final}
\end{equation}
where $\psi$ is periodic coordinate and $-\infty< T <\infty$.

Consider an azimuthal curve $\gamma$ defined by $r= r_0, z=z_0, T=T_0$, where $r_0>0$, $z_0$, and $T_0$ are constants. From (\ref{final}), we have  
\begin{equation}
ds^2=-T\,\sinh^{2} r\,d\psi^2.
\label{reduce}
\end{equation}
These curves are null (or null geodesics) at $T=T_0=0$, spacelike throughout $T=T_0<0$, but become timelike for $T=T_0>0$, which indicates the presence of CTCs. Hence, CTCs form at a definite instant of time satisfying $T=T_0>0$. The above analysis is valid provided that the CTCs evolve from an initially spacelike $T=const$ hypersurface \cite{Ori2}. A hypersurface $T=const$ is spacelike provided $g^{TT}<0$ for $T<0$, timelike provided $g^{TT}>0$ for $T>0$, and null hypersurface provided $g^{TT}=0$ for $T=0$. From metric (\ref{final}), we found that  
\begin{equation}
g^{TT}=\frac{T}{\sinh^{2} r}.
\label{norm-space}
\end{equation}
Thus, the hypersurface $T=const=T_0<0$ is spacelike ($r\neq 0$) and can be chosen as initial hypersurface over which the initial data may be specified. There is a Cauchy horizon at $T=T_0=0$ for any such initial hypersurface $T=T_0<0$. Hence, the spacetime evolves from an initial spacelike hypersurface in a causally well-behaved manner, up to a moment, that is, a null hypersurface $T=T_0=0$, and the formation of CTC takes place from causally well-behaved initial conditions. Assuming that the evolution beyond the Cauchy horizon proceeds in an analytic manner, we recover the region $T>0$ as well, and CTCs appear.

The formation of CTCs is thus identical to the Misner space. The Misner space is a two-dimensional locally flat metric with peculiar identifications. The metric of Misner space in 2D \cite{Mis} is
\begin{equation}
ds^{2}_{Misn}=-2\,dT\,dX-T\,dX^2,
\label{misner}
\end{equation}
where $-\infty< T <\infty$ but the coordinate $X$ is periodic. The curves defined by $T=T_0>0$, where $T_0$ is a constant, are timelike and closed (since $X$ is periodic), thus CTCs formed. Note that the metric of Misner space (\ref{misner}) can be obtained from the Minkowski metric
\begin{equation}
ds^{2}_{Mink}=-dt^2+dx^2,
\label{Mink}
\end{equation} 
by applying a similar transformation (\ref{trans}).

\section{Conclusion} 

In this paper, we present a cylindrically symmetric solution of the field equations which is regular everywhere except on the symmetry axis where it possesses a naked curvature singularity. The spacetime is of type D in the Petrov classification and is asymptotically flat radially and its matter-energy content is an anisotropic fluid that satisfies the different energy conditions. The nonzero kinematic parameters associated with the fluid four-velocity vector, namely, the acceleration diverges on the symmetry axis and vanishes at large radial distance. Similarly, the physical parameters, the energy density, the radial pressure and the tangential pressure of the anisotropic fluid, behave the same way. Finally, we generated a rotating spacetime admitting CTCs and these curves evolve from an initial spacelike hypersurface in a causally well behaved manner. The possibility that a naked curvature singularity gives rise to a Cosmic Time Machine has been discussed by Clarke and his collaborator \cite{Clark}. The presented cylindrically symmetric spacetime may represent such a Cosmic Time Machine.

Competing Interests : The author declares that there are no competing interests regarding publication of this paper.

\end{document}